\documentclass[prb,twocolumn,showpacs,preprintnumbers,amsmath,amssymb,floatfix,superscriptaddress,10pt,aps]{revtex4-1}

\usepackage{graphicx,color,hyperref}

\begin{document}

\title{Stability of trions in coupled quantum wells modelled by
  two-dimensional bilayers}

\author{O.\ Witham}

\affiliation{Institut f{\"u}r Theoretische Physik,
  Goethe-Universit{\"a}t Frankfurt, 60438 Frankfurt am Main, Germany}

\affiliation{Department of Physics, Lancaster University, Lancaster
  LA1 4YB, United Kingdom}

\author{R.\ J.\ Hunt}

\affiliation{Department of Physics, Lancaster University, Lancaster
  LA1 4YB, United Kingdom}

\author{N.\ D.\ Drummond}

\affiliation{Department of Physics, Lancaster University, Lancaster
  LA1 4YB, United Kingdom}

\date{\today}

\begin{abstract}
We report variational and diffusion quantum Monte Carlo calculations
of the binding energies of indirect trions and biexcitons in ideal
two-dimensional bilayer systems within the effective-mass
approximation, and with a Coulomb $1/r$ interaction between charge
carriers.  The critical layer separation at which trions become
unbound has been studied for various electron-hole mass ratios, and
found to be over an order of magnitude larger than the critical layer
separation for biexcitons.
\end{abstract}


\maketitle

\section{Introduction \label{sec:intro}}

Excitonic complexes consisting of bound states of small numbers of
electrons and holes have been observed under many different conditions
in semiconductors. In the dilute limit, excitons collectively behave
as a gas of weakly interacting neutral bosons. It was first predicted
in the 1960s that Bose-Einstein condensates of excitons might form
under experimentally accessible
conditions,\cite{Blatt1962,Keldysh1968} and coupled quantum-well
heterostructures, in which electrons and holes are confined to
spatially separated layers, were later identified as an ideal testbed
for the observation of an excitonic
condensate.\cite{Lozovik1975,Zhu1995} Such geometries hinder
recombination and significantly increase the exciton lifetime, which
is necessary if thermalization of a photoexcited gas of excitons is to
occur.  Furthermore, at sufficiently large layer separations, the
repulsive dipole-dipole interaction between spatially indirect
excitons helps to prevent the formation of larger charge-carrier
complexes.

A Bose-Einstein condensate of excitons in a semiconductor such as GaAs
is expected to be dark, i.e., not to couple directly to
photons.\cite{Combescot2007} Striking fragmented-ring patterns of
indirect exciton photoluminescence have been observed around laser
excitation spots in GaAs coupled quantum
wells,\cite{Butov2002,Snoke2002} and subsequent work\cite{Yang2006}
has indicated that these patterns arise due to indirect excitons
travelling outwards in a dark, coherent state from the point at which
they are created.  More recent experimental work has provided further
evidence for the creation of an excitonic Bose-Einstein condensate in
coupled quantum wells, as revealed by the macroscopic spatial
coherence of indirect excitons at low
temperatures.\cite{Timofeev2008,High2012,Shilo2013,Alloing2014}
However, despite the many exciting and important advances in this
field, the situation is not perfectly clearcut and there remains a
need to analyze factors that could influence the formation of an
excitonic condensate in a coupled quantum well.\cite{Combescot2017}

The formation of indirect biexcitons (also neutral composite bosons,
but with a higher mass than excitons) could potentially inhibit
condensation, but various theoretical
works\cite{Zimmermann2007,Schindler2008,Meyertholen2008,Lee2009}
showed that biexcitons are expected to be unbound in the typical
coupled quantum-well geometries accessible to experimentalists, in
which the layer separation is of the order of tens of
{\AA}ngstr\"{o}ms.  In this work we investigate the stability of a
different class of charge-carrier complex, namely indirect trions
(bound states of two electrons in one layer and a hole in the opposite
layer, or \textit{vice versa}) in coupled quantum wells.  Trion
formation can occur if a nonzero concentration of free charge carriers
is present in a semiconductor, or if the different mobilities of
electrons and holes leads to a local imbalance in the carrier
concentration. Free charge carriers can then bind to excitons created
by optical excitation, producing trions.  A finite concentration of
trions effectively provides disorder, and could therefore tend to
restrict Bose-Einstein condensation of the remaining excitons.  As an
additional motivation for our work, it is known that trions play a key
role in the optical properties of atomically thin transition-metal
dichalcogenides and other two-dimensional (2D)
semiconductors.\cite{Mak2013,Zhang2014,Srivastava2015,Zhang2015,Jones2013,Szyniszewski2017}
We shall assume an isotropically screened Coulomb interaction, as
appropriate for trions in GaAs/AlGaAs heterostructures, rather than
the Keldysh interaction appropriate for charge carriers in 2D systems
with a significant in-plane
polarizability.\cite{Keldysh1979,Mostaani2017} Nevertheless, our
results are of qualitative relevance to optical studies of bilayers of
2D semiconductors.

\begin{figure}[!htbp]
\begin{center}
\includegraphics[clip,width=\linewidth]{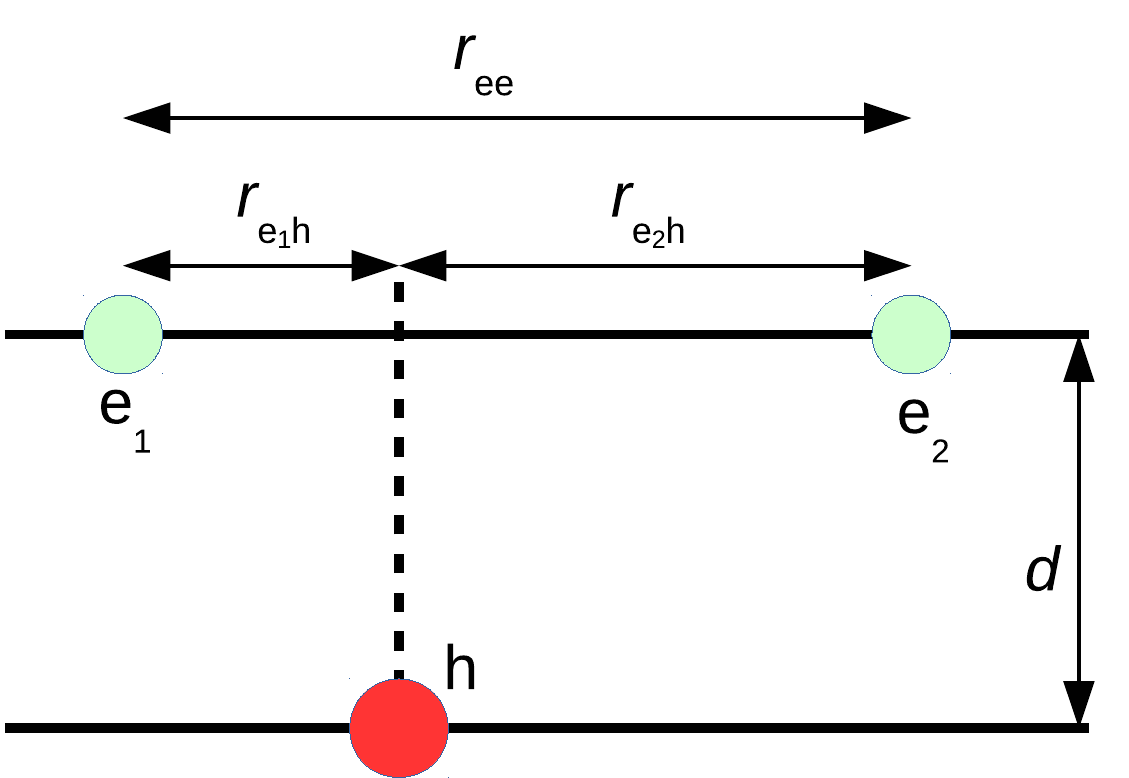}
\caption{(Color online) Schematic arrangement of two electrons (light
  green) and a hole (red) for a negative trion in a coupled quantum
  well, showing the definition of the interparticle distances. The two
  electrons are in a spin singlet state and hence act as
  distinguishable particles in the spatial wave function.  The
  electrons and hole move in 2D in spatially separated
  layers. \label{fig:trion_schematic}}
\end{center}
\end{figure}

We have used the variational and diffusion quantum Monte Carlo (VMC
and DMC) methods\cite{Ceperley1980,Foulkes2001} to calculate the
ground-state energies and binding energies of spatially indirect
trions within the 2D-isotropic effective-mass approximation. A
negative trion in such an ideal 2D bilayer system is approximately
described by the Hamiltonian
\begin{eqnarray}
\hat{H} & = & -\frac{\hbar^2}{2m_\text{e}}\nabla_{\text{e}_1}^2
-\frac{\hbar^2}{2m_\text{e}}\nabla_{\text{e}_2}^2-
\frac{\hbar^2}{2m_\text{h}}\nabla_\text{h}^2 +\frac{e^2}{4\pi\epsilon
  r_{\text{ee}}} \nonumber \\ & & ~~~{} -\frac{e^2}{4\pi\epsilon
  \sqrt{r_{\text{e}_1 \text{h}}^2+d^2}}- \frac{e^2}{4\pi\epsilon
  \sqrt{r_{\text{e}_2\text{h}}^2+d^2}},
\label{eq:gen_Ham}
\end{eqnarray}
where $m_\text{e}$ and $m_\text{h}$ are the electron and hole
effective masses, $e$ is the magnitude of electronic charge, $d$ is
the interlayer separation, and $\epsilon$ is the permittivity of the
medium in which the two layers are embedded.  The in-plane
interparticle distances are given by $r_\text{ee}=|{\bf
  r}_{\text{e}_1}-{\bf r}_ {\text{e}_2}|$ and
$r_{\text{e}_i\text{h}}=|{\bf r}_{\text{e}_i}-{\bf r}_\text{h}|$,
where ${\bf r}_{\text{e}_1}$, ${\bf r}_{\text{e}_2}$, and ${\bf
  r}_\text{h}$ are 2D vectors holding the in-plane coordinates of the
electrons and the hole, which are assumed to be confined to parallel
planes as shown in Fig.\ \ref{fig:trion_schematic}. Although our
results pertain directly to the negative trion (X$^-$), the
corresponding properties of the positive trion (X$^+$) can easily be
generated by charge conjugation, i.e., by interchanging $m_\text{e}$
and $m_\text{h}$ (or, equivalently, by replacing $\sigma$ by
$\sigma^{-1}$).  Furthermore, we consider only the ground-state case
in which the two electrons are distinguishable (opposite-spin
electrons). Trions and biexcitons with indistinguishable electrons are
much less stable than trions with distinguishable electrons; by
analogy with results obtained for biexcitons in single-layer 2D
semiconductors, we expect that trions and biexcitons with
indistinguishable particles are only stable when the indistinguishable
particles are very heavy, so that exchange effects are
negligible.\cite{Mostaani2017}

Although biexcitons in coupled quantum wells have been studied
extensively, there are relatively few theoretical studies of indirect
trions. Kulakovskii and Lozovik\cite{Kulakovskii2002} and Berman
\textit{et al.}\cite{Berman2014}\ study trions consisting of a direct
exciton in one layer bound to a charge carrier in a neighboring layer,
and make the interesting suggestion that Wigner crystallization of
trions could occur.  Kovalev and Chaplik\cite{Kovalev2008}
investigated the behavior of indirect trions in an out-of-plane
magnetic field by treating the interlayer electron-hole Coulomb
potential within a harmonic approximation.  Sergeev and
Suris\cite{Sergeev2003} studied a model similar to ours within a
heavy-hole approximation.  We compare our results with theirs in
Sec.\ \ref{sec:trion_reuslts}.

The rest of this paper proceeds as follows. In
Sec.\ \ref{sec:methodology} we describe our methodology for
calculating the binding energies of trions and biexcitons; in
Sec.\ \ref{sec:results} we present our results; and finally in
Sec.\ \ref{sec:conclusions} we draw our conclusions.  Except where
otherwise stated, we use excitonic units: energies are given in terms
of the exciton Rydberg $R_\text{y}^*=\mu e^4/[2{(4\pi \epsilon
    \hbar)}^2]$ and lengths are given in terms of the exciton Bohr
radius $a_0^*=4 \pi\epsilon \hbar^2/ (\mu e^2)$, where $\mu=m_\text{e}
m_\text{h}/(m_\text{e} + m_\text{h})$ is the reduced mass of the
electron and hole.  In these units the dimensionless solutions
$E/R_\text{y}^*$ to the Schr\"{o}dinger equation only depend on the
electron-hole mass ratio $\sigma=m_\text{e}/m_\text{h}$ and the
dimensionless layer separation $d/a_0^*$.

\section{Computational methodology \label{sec:methodology}}

The VMC and DMC methods\cite{Ceperley1980,Foulkes2001} as implemented
in the \textsc{casino} code\cite{Needs2010} were used in conjunction
with a trial wave function of the form
\begin{equation}
\Psi = \exp(J) \Psi_\text{ee} \Psi_{\text{e}_1\text{h}}
\Psi_{\text{e}_2\text{h}}, \label{eq:final_wf} \end{equation} where
\begin{eqnarray}
\Psi_\text{ee} & = & \exp\left[ \frac{c_1r_{\text{ee}}}{1+c_2
    r_{\text{ee}}}+\frac{1}{2}\left(e^{-c_6
    r_{\text{ee}}^2}-1\right)\log(r_{\text{ee}})\right] \label{eq:psi_ee}
\\ \Psi_{\text{e}_1\text{h}} & = & \exp\left[ \frac{c_3
    r_{\text{e}_1\text{h}}+c_4 r_{\text{e}_1\text{h}}^2}{1+c_5
    r_{\text{e}_1\text{h}}}\right] \\ \Psi_{\text{e}_2\text{h}} & = & \exp \left[
  \frac{c_3 r_{\text{e}_2\text{h}}+c_4 r_{\text{e}_2\text{h}}^2}{1+c_5 r_{\text{e}_2\text{h}}}\right], \label{eq:psi_e2h}
\end{eqnarray}
to perform total-energy calculations for the negative trion. Our trial
wave function is similar to the one used by Tan \textit{et al.}\ to
study biexcitons,\cite{Tan2005} but with an additional term that
describes the behavior of the wave function when one electron is far
from the remaining exciton. The nested exponential in
Eq.\ (\ref{eq:psi_ee}) acts as a ``switching on'' term; the logarithm
only manifests appreciably in the case that the two electrons are far
apart.  The Jastrow exponent $J$ contains cuspless two-body and
three-body polynomial functions of the interparticle distances,
truncated at finite range.\cite{Drummond2004,Lopez2012} The wave
function of Eqs.\ (\ref{eq:final_wf})--(\ref{eq:psi_e2h}) incorporates
both short- and long-range effects through the use of the
Pad\'{e}-form exponents. It exhibits the correct symmetry of the
ground-state system, being invariant upon the exchange of the two
electrons, i.e., $\Psi({\bf r}_{\text{e}_1},{\bf r}_{\text{e}_2},{\bf
  r}_\text{h})=\Psi({\bf r}_{\text{e}_2},{\bf r}_{\text{e}_1}, {\bf
  r}_\text{h})$. The trial wave function reduces to
\begin{equation}
\Psi\to A\frac{1}{\sqrt{r_{\text{ee}}}}\exp[-k
r_{\text{ee}}]\exp\left[\frac{c_3
  r_{\text{e}_2\text{h}}+c_4 r_{\text{e}_2\text{h}}^2}{1+c_5
  r_{\text{e}_2\text{h}}}\right], \label{eq:psi_ee_lr}
\end{equation}
when one electron is far from the remaining exciton, where $k>0$ and
$A$ is constant. This form of wave function is appropriate for an
electron moving in the potential energy due to the static charge
distribution of the remaining indirect exciton. The static dipole
moment of an exciton is ${\bf p}=-ed{\bf e}_z$, where ${\bf e}_z$ is a
unit vector in the $z$ direction. Let the separation of the electron
from the center of the exciton be ${\bf r}+(d/2){\bf e}_z$, where
${\bf r}$ is the in-plane separation.  Hence the long-range
exciton-electron interaction energy is
\begin{equation} -e{\bf p}\cdot[{\bf
    r}+(d/2){\bf e}_z]/\left[4\pi\epsilon|{\bf r}+(d/2){\bf
      e}_z|^3\right] \sim r^{-3}. \end{equation} Solving the 2D radial
Schr\"{o}dinger equation for motion in a rapidly decaying potential
such as $r^{-3}$ gives the form of wave function shown in
Eq.\ (\ref{eq:psi_ee_lr}) at long range.

The parameter set $\{c_1,\ldots,c_6\}$ is subject to the following
conditions: (i) The values of $c_1$ and $c_3$ are fixed by the
electron-electron and electron-hole Kato cusp
conditions;\cite{Kato1957,Pack1966} (ii) $c_4<0$ to ensure that the
wave function falls off exponentially as $r_{\text{e}_i\text{h}}
\rightarrow \infty$; (iii) $c_2,c_5>0$ to avoid divergences in the
wave function; and (iv) $c_6>0$ to enforce the correct long-range
behavior of an electron in a dipole field.  The optimal values of the
parameters $\{c_1,\ldots,c_6\}$ were obtained by successive
minimization of the variance of the local energy and the energy
expectation value.\cite{Umrigar1988,Drummond2005,Toulouse2007}

In the DMC method the ground-state component of the trial wave
function is projected out by simulating a stochastic process governed
by the Schr\"{o}dinger equation in imaginary time.  In systems such as
those considered here, in which there are no indistinguishable
fermions, there are no uncontrolled approximations in the DMC
ground-state energy.  We simultaneously remove time-step bias and
population-control bias by performing DMC calculations at two
different, small time steps, with the walker population being in
inverse proportion to the time step, and extrapolating linearly to
zero time step.

The binding energy $E_{\text{X}^-}^\text{b}$ of a negative trion is
defined as the energy required to split the trion into an exciton and
a free electron, i.e.,
\begin{equation}
  E_{\text{X}^-}^\text{b} = E_{\text{X}} - E_{\text{X}^-},
\end{equation}
where $E_\text{X}$ and $E_{\text{X}^-}$ are the ground-state total
energies of an exciton and a negative trion, respectively. Instability
of the trion with respect to dissociation into a free electron and an
indirect exciton is signalled by difficulty optimizing a
bound-state trial wave function, followed by the occurrence of
nonpositive $E_{\text{X}^-}^{\text{b}}$ values in DMC calculations in
which the trial wave function is forced to be bound. The curve defined
by $E_{\text{X}^-}^{\text{b}}(d/a_0^*,\sigma) = 0$, where
$E_{\text{X}^-}^{\text{b}}(d/a_0^*,\sigma)$ is the binding energy of
the trion with a given electron-hole mass ratio $\sigma$ and
dimensionless layer separation $d/a_0^*$, defines the boundary of the
stability region of the trion. In practice we invert this relation
and simply quote the critical layer separation
$d_{\text{X}^-}^\text{crit}(\sigma)$ as a function of mass ratio. We
have attempted to probe the trion stability region directly by
studying systems with electron-hole mass ratios $\sigma=1/4$, $1/2$,
$3/4$, $1$, $4/3$, $2$, and $4$, and fitting the resultant trion
binding energies to a Pad\'{e} approximant of the form
\begin{equation}
  \frac{E_{\text{X}^-}^\text{b}(d/a_0^*)}{R_\text{y}^*} =
  \frac{E_{\text{X}^-}^\text{b}(0)/R_\text{y}^* + \sum^3_{i=1} a_i (d/a_0^*)^i}{1 +
    \sum^4_{j=1} b_j (d/a_0^*)^j}, \label{eq:trion_BE_fixedsigma}
\end{equation}
where $\{a_i\}$ and $\{b_j\}$ are fitting parameters.  We have found
that this functional form yields sufficiently accurate fits of the
binding energies for all mass ratios considered here.

Where the trion is bound, we have performed fits to a ``partial 2D''
Pad\'{e} approximant of the form
\begin{equation}
  \frac{E_{\text{X}^-}^\text{b}(\sigma, d/a_0^*)}{R_\text{y}^*} =
  \frac{\sum^3_{i=0} \sum^3_{j=0} f_{ij}(1 +
    \sigma)^{-i/2}(d/a_0^*)^j} {1 + \sum^4_{k=1} g_k
    (d/a_0^*)^{k}},\label{eq:2d_pade}
\end{equation}
where $f_{ij}$ and $g_k$ are fitting parameters, and the $\sigma$
dependence is motivated by the harmonic approximation within the
Born-Oppenheimer approximation in the case that $\sigma \rightarrow
\infty$.\cite{Sergeev2003,Spink2016,Mostaani2017}

Finally, we report pair-distribution functions (PDFs), which give
information about the structure and spatial extent of charge-carrier
complexes.  For a negative trion, the electron-electron and
electron-hole PDFs are defined via
\begin{eqnarray} g_{\text{X}^{-}}^\text{ee}({\bf r}) & = & \left< \delta \left({\bf r}-{\bf r}_\text{ee}\right) \right> \\ g_{\text{X}^{-}}^\text{eh}({\bf r}) & = &
  \left< \sum_{i=1}^2 \delta\left({\bf r}-{\bf
    r}_{\text{e}_i\text{h}}\right) \right>. \end{eqnarray} The PDFs in
a biexciton are defined in an analogous fashion. The PDFs are
evaluated by binning the interparticle distances sampled in the VMC and
DMC calculations.  The error in the VMC and DMC estimates of the PDF is
first order in the error in the trial wave function; however, the
error in the extrapolated estimate of the PDF, given by two times the
DMC result minus the VMC result, is second order in the error in the
trial wave function.\cite{Ceperley1979} The PDF results that we report
were obtained by extrapolated estimation.

\section{Results and discussion \label{sec:results}}

\subsection{Trion binding energy and region of stability
\label{sec:trion_reuslts}}

\begin{figure}[!htbp]
\begin{center}
\includegraphics[clip,width=\linewidth]{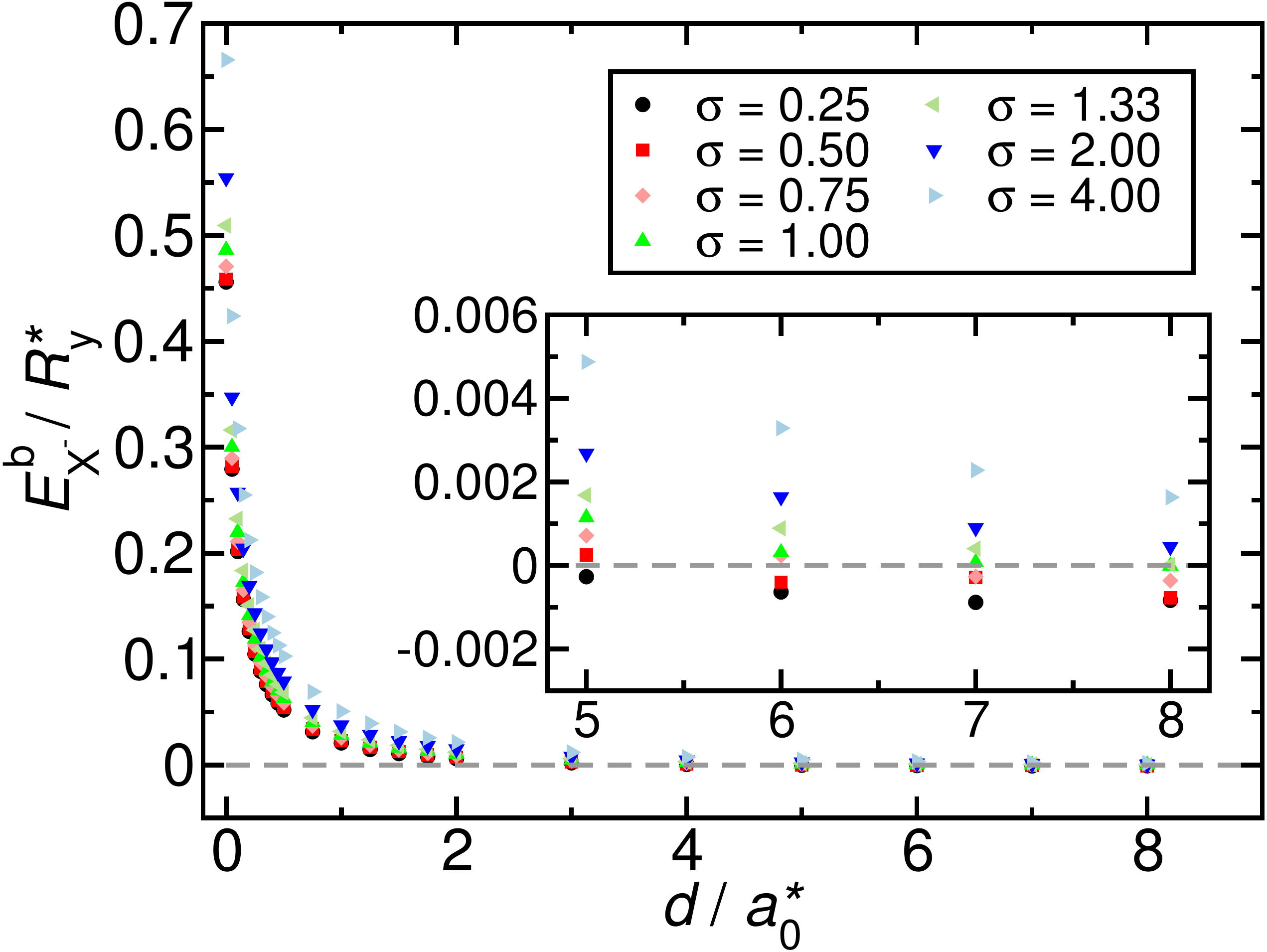}
\caption{(Color online) Negative-trion binding energy as a function of
  interlayer spacing $d$ and electron-hole mass ratio $\sigma$, in
  excitonic units. The inset shows the edge of the region of stability
  for negative trions in greater
  detail. \label{fig:tri_binding_energies}}
\end{center}
\end{figure}

Our negative-trion binding-energy results are displayed in
Fig.\ \ref{fig:tri_binding_energies}, and our predicted critical layer
separations are shown in Table \ref{table:critical_separation}.  For
larger electron-hole mass ratios $\sigma$, the binding energy decays
slowly to zero, and furthermore the DMC calculations become more
difficult due to the separation of imaginary-time scales for the
different particles.  The inset to
Fig.\ \ref{fig:tri_binding_energies} illustrates the challenge in
determining a precise value for the boundary of the region of
stability.  Nevertheless, we can easily place lower bounds on the
critical layer separation at a given mass ratio.  As can be seen in
Fig.\ \ref{fig:tri_binding_energies} and Table
\ref{table:critical_separation}, trions are stable over a large region
of the $(d/a_0^*,\sigma)$ model parameter space.  When compared with
the biexciton stability region,\cite{Lee2009} we find that the trion
is bound for layer separations over an order of magnitude larger than
those for which the biexciton is bound.  Trion formation is always
possible whenever biexcitons are bound, and trion binding energies are
typically far larger for the same set of material parameters.

\begin{table}[!htbp]
\caption{Critical layer separations for negative trions
  ($d_{\text{X}^-}^\text{crit}$) and biexcitons
  ($d_\text{XX}^\text{crit}$) at different electron-hole mass ratios
  $\sigma$.  The biexciton critical layer separations were evaluated
  using Eq.\ (2) of
  Ref.\ \onlinecite{Lee2009}. \label{table:critical_separation}}
\begin{center}
\begin{tabular}{lcc}
\hline \hline

$\sigma$ & $d_{\text{X}^-}^\text{crit}/a_0^*$ &
$d_\text{XX}^\text{crit}/a_0^*$ (Ref.\ \onlinecite{Lee2009}) \\

\hline

$1/4$ & $4.52(4)$ & $0.48$ \\

$1/2$ & $5.26(6)$ & $0.43$ \\

$3/4$ & $6.63(9)$ & $0.42$ \\

$1$ & $7.69(7)$ & $0.42$ \\

$4/3$ & $>8$ & $0.42$ \\

$2$ & $>8$ & $0.43$ \\

$4$ & $>8$ & $0.48$ \\

\hline \hline
\end{tabular}
\end{center}
\end{table}

The biexciton (XX), whose dominant decay is into a pair of excitons
($\text{XX} \rightarrow \text{X} + \text{X}$), has a stability region
that is determined by the effective interactions of the constituent
indirect excitons, and this effective interaction is a repulsive
dipole-dipole interaction at long range.\cite{Schindler2008,Lee2009}
Negative-trion dissociation is determined by the effective interaction
of a lone electron with a single indirect exciton.  The interaction
potential between an indirect exciton and a lone electron consists of
a repulsive part due to the static charge distribution of the exciton,
which falls off as $r^{-3}$, and an attractive part due to the induced
dipole moment of the exciton, which falls off as $r^{-4}$.  Over the
intermediate range the attractive part of the interaction plays a much
more important role in the trion than in the biexciton.

Fitting Eq.\ (\ref{eq:2d_pade}) to our trion binding energies results
in a maximum error of $5 \times 10^{-4} R_\text{y}^*$, with over 90\%
of the data points falling within $2 \times 10^{-4} R_\text{y}^*$ of the
fit.  The fitting parameters $f_{ij}$ and $g_k$ are\footnote{The
  top-left element of the matrix $f$ in Eq.\ (\ref{eq:fmatrix}) is the
  fitting parameter $f_{00}$, and so on.}
\begin{eqnarray}
f & = & \begin{pmatrix}
\begin{array}{cccc}
 1.408  &  21.53  &  25.25  &  1.676  \\
-2.340  & -40.43  & -36.22  & -11.51  \\
 1.617  &  30.47  &  5.803  &  17.36  \\
-0.2129 & -0.5492 &  7.423  & -8.694
\end{array}
\end{pmatrix} \label{eq:fmatrix} \\
{\bf g} & = & \begin{pmatrix}
26.16 \\
147.7 \\
186.4 \\
29.45
\end{pmatrix}.
\end{eqnarray}
This fit applies only for $1/4 \leq \sigma \leq 4$ and $0 \leq d \leq
\min \{8a_0^*,d_{\text{X}^-}^\text{crit}\}$. Accurately parameterizing
the binding energy near the critical separation
$d_{\text{X}^-}^\text{crit}$ is not possible with our limited data,
and caution should be applied when relying on this fit near the
critical region.

Appropriate model parameters for the GaAs/AlGaAs coupled quantum-well
device studied by Butov \textit{et al.}\cite{Butov2002}\ were
identified in Ref.\ \onlinecite{Tan2005}.  In particular the layer
separation $d$ was chosen such that the exciton binding energy
obtained using the screened Coulomb interaction with strictly 2D
electrons and holes matches the exciton binding energy obtained using
a more realistic model for electrons and holes moving in the quantum
wells.\cite{Szymanska2003} The electron and hole masses are taken to
be $m_\text{e}=0.067m_0$ and $m_\text{h}=0.134m_0$, where $m_0$ is the
bare electron mass.  The permittivity is taken to be
$\epsilon=13.2\epsilon_0$.  Hence the mass ratio is $\sigma=0.5$, the
exciton Bohr radius is $a_0^*=156$ {\AA}, and the exciton Rydberg is
$R_\text{y}^*=3.5$ meV\@.  Finally, the layer separation is taken to
be $d=100$ {\AA}${}=0.64a_0^*$.

With these parameters, Lee \textit{et al.}\cite{Lee2009}\ found the
critical layer separation for biexciton formation to be
$d_\text{XX}^\text{crit}(0.5) = 0.43(5)a_0^* = 67(8)$ {\AA}. This is
significantly less than the actual layer separation, implying that
biexcitons are unbound. On the other hand, we find the critical layer
separation for negative-trion formation to be
$d_{\text{X}^-}^\text{crit}(0.5) = 5.26(6)a_0^* = 821(9)$ {\AA} and
the critical layer separation for positive-trion formation to be
$d_{\text{X}^-}^\text{crit}(2) >8a_0^* = 1248$ {\AA}.  Both of these
are many times larger than the actual layer separation, implying that
both positive and negative trions are bound.  Using
Eq.\ (\ref{eq:trion_BE_fixedsigma}), the predicted binding energies of
negative and positive trions are $0.0411(4)R_\text{y}^*=0.14$ meV and
$0.06166(3)R_\text{y}^*=0.22$ meV, respectively.  Hence both positive
and negative trions are expected to be present at temperatures below
$T=2$ K, which corresponds to an energy of about $k_\text{B}T=0.17$
meV\@.

Our results show that isolated trions are bound in realistic
coupled-quantum-well geometries.  By continuity, we expect trions to
persist as quasiparticles at low, finite charge-carrier concentrations
in these systems.  A well-defined direct trion persists in a host
monolayer 2D electron gas at low or intermediate electron
concentrations, up to the point where the density parameter $r_{\rm
  s}$ is about three times the exciton Bohr radius.\cite{Spink2016}
Screening of the electron-hole attraction by a finite concentration of
charge carriers in an electron-hole bilayer is expected to reduce the
binding energy of an indirect trion and hence to reduce the critical
layer separation relative to that of an isolated trion.

Sergeev and Suris have studied indirect trions in coupled quantum
wells using the same model as us, but making use of a heavy-hole
approximation.\cite{Sergeev2003} They used a variational approach to
find an approximate solution to the 2D Schr\"{o}dinger equation of a
single electron in one layer moving in the potential supplied by two
fixed holes in the other layer.  The resulting ground-state energy as
a function of hole separation provided a Born-Oppenheimer
potential-energy surface for the two holes. They then numerically
solved the Schr\"{o}dinger equation for the relative motion of the
holes in that potential-energy surface to predict X$^+$ binding
energies in GaAs and ZnSe coupled quantum wells. In the case of GaAs,
at a layer separation of $d=0.42a_0^*$, Sergeev and Suris predict an
X$^+$ binding energy of 0.43 meV, whereas the fit to our DMC results
(after charge conjugation) yields 0.63 meV\@. In the case of ZnSe, at
$d=1.25a_0^*$, Sergeev and Suris predict a binding energy of 0.48 meV;
however fits to our DMC results yield 0.77 meV\@. In order to make these
comparisons as fair as possible, we have taken excitonic Rydberg units
and electron-hole mass ratios identical to Sergeev and Suris (for
GaAs, $R_\text{y}^*=4.84$ meV and $\sigma=0.196$; for ZnSe,
$R_\text{y}^*=20$ meV and $\sigma=0.26$).

\subsection{Biexciton binding energy}

In an extension to the earlier work of Lee \textit{et
  al.},\cite{Lee2009} for completeness we provide an accurate
parameterization of the biexciton binding energy in the bound region.
We have used the same trial wave function form as in
Ref.\ \onlinecite{Tan2005}, multiplied by a polynomial Jastrow
factor.\cite{Drummond2004,Lopez2012} We have fitted the function
\begin{equation}
\frac{E_\text{XX}^{\text{b}}(\sigma, d/a_0^*)}{R_\text{y}^*} =
\frac{\sum^2_{i=0} \sum^2_{j=0} F_{ij} {\left(\sigma +
      \sigma^{-1} \right)}^{i/2}(d/a_0^*)^{j}} {\sum^{3}_{k=0}G_k
  (d/a_0^*)^{k}} \label{eq:biex_binding_fit}
\end{equation}
to our biexciton binding-energy data, where $F_{ij}$ and $G_{k}$ are
fitting parameters.  This obeys the necessary symmetry under charge
conjugation ($\sigma \rightarrow \sigma^{-1}$). As in
Eq.\ (\ref{eq:2d_pade}), the square-root behavior in
Eq.\ (\ref{eq:biex_binding_fit}) arises from harmonic motion in the
Born-Oppenheimer approximation.\cite{Mostaani2017} Our fit of
Eq.\ (\ref{eq:biex_binding_fit}) has a maximum error of $2 \times
10^{-2} R_\text{y}^*$, with over 90\% of the data points falling
within $5 \times 10^{-3} R_\text{y}^*$ of the fit.  The results of
this fit are
\begin{eqnarray}
F & = & \begin{pmatrix}
\begin{array}{ccc}
 0.03495 & -0.9822 & 2.437 \\
 0.07670 & 0.3303 & -1.786 \\
-0.005277 & -0.02931 & 0.2942
\end{array}
\end{pmatrix} \\
{\bf G} & = & \begin{pmatrix}
0.1726\\
3.256\\
1.567\\
29.95
\end{pmatrix}.
\end{eqnarray}

\subsection{PDFs}

\begin{figure}[!htbp]
\begin{center}
\includegraphics[clip,width=\linewidth]{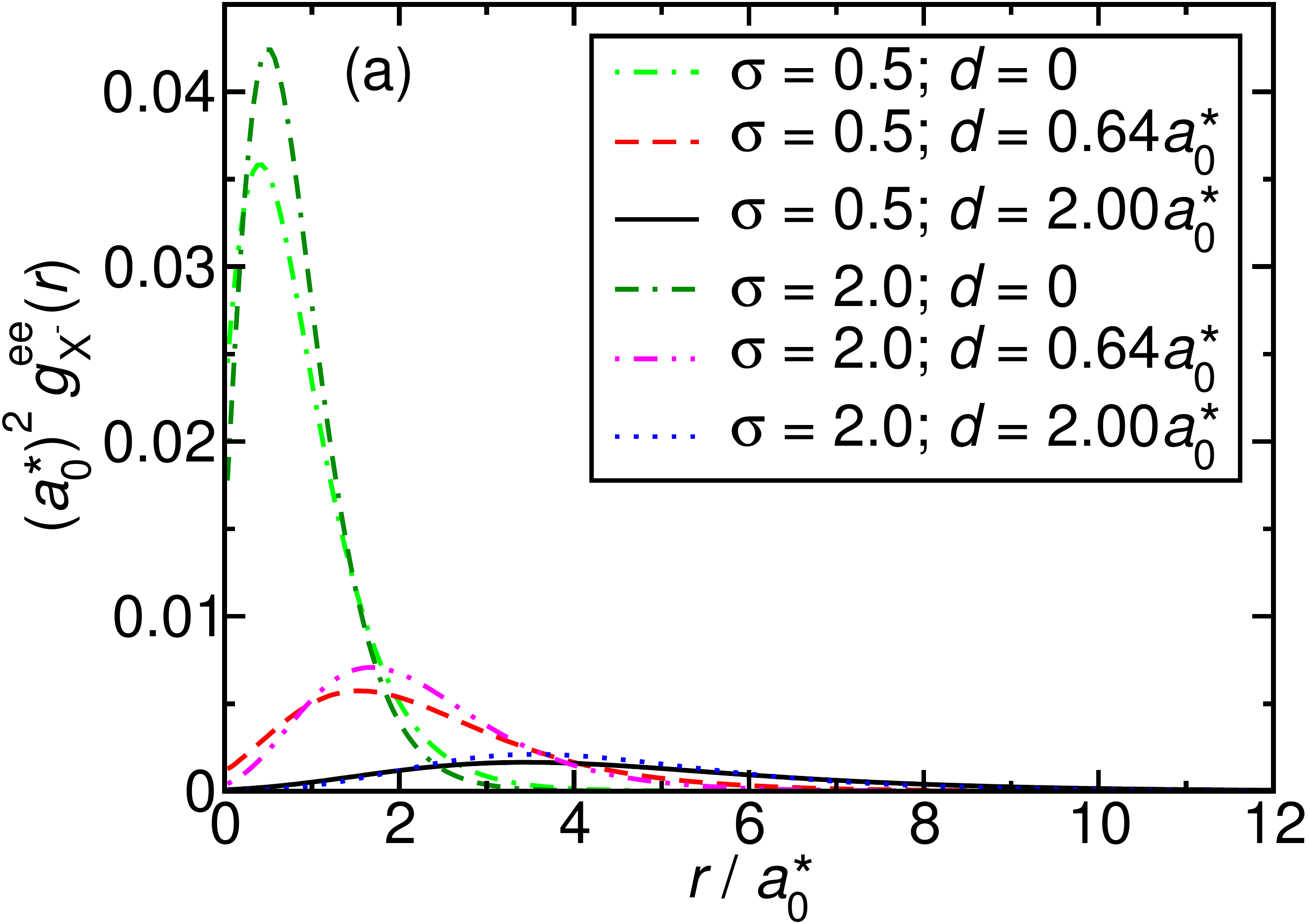} \\[1ex]
\includegraphics[clip,width=\linewidth]{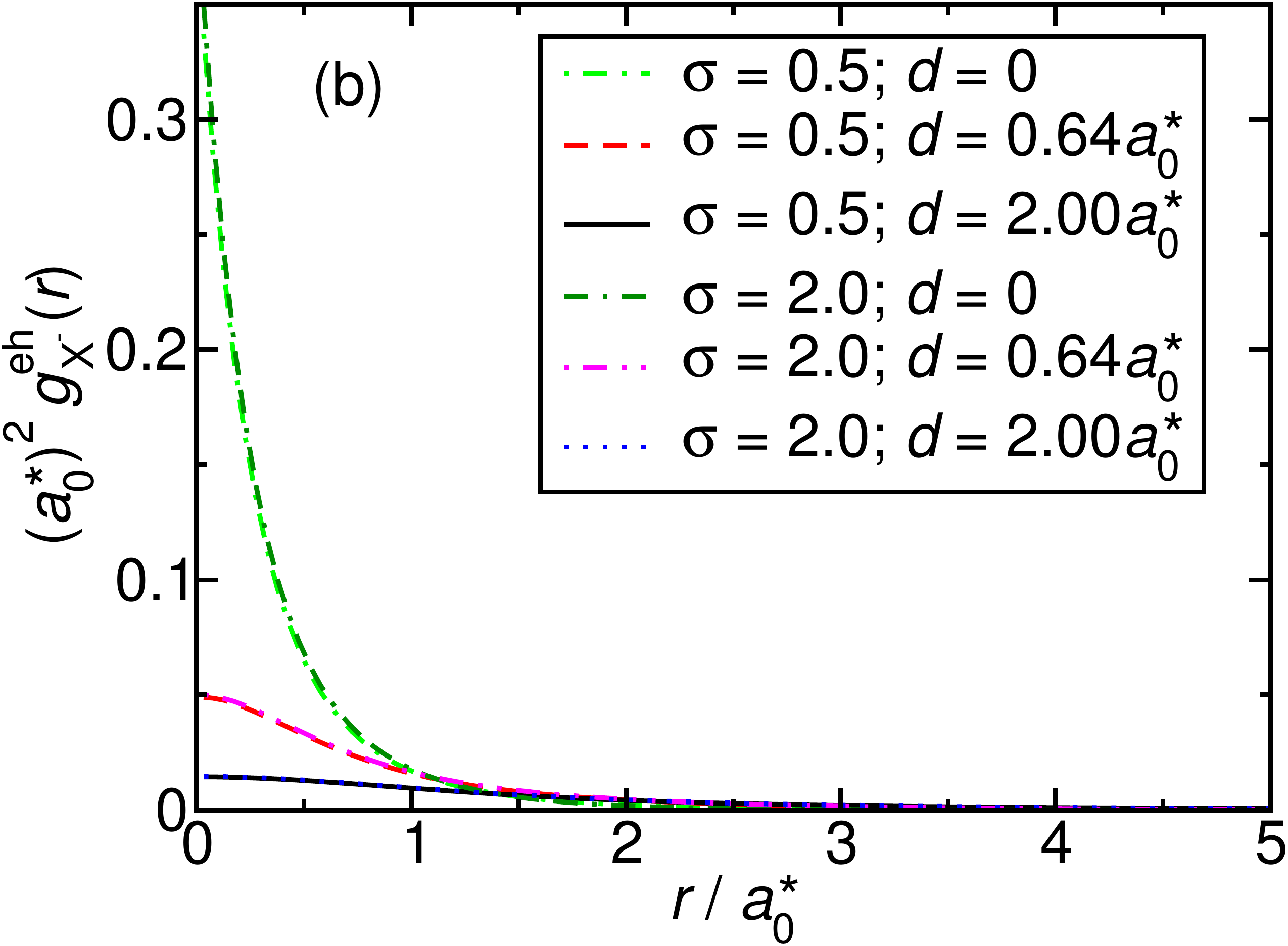}
\caption{(Color online) (a) Electron-electron
  [$g_{\text{X}^-}^\text{ee}(r)$] and (b) electron-hole
  [$g_{\text{X}^-}^\text{eh}(r)$] pair-distribution functions for
  indirect negative trions in bilayers with different separations $d$
  and electron-hole mass ratios $\sigma$. The $x$-axis shows the
  in-plane separation.  \label{fig:g}}
\end{center}
\end{figure}

In Fig.\ \ref{fig:g} we plot electron-electron and electron-hole
pair-distribution functions for negative trions.  It is clear that the
spatial extent of the trion increases rapidly with layer separation,
but is relatively insensitive to mass ratio.

\section{Conclusions \label{sec:conclusions}}

We have generated statistically exact total-energy data for indirect
trions and biexcitons in a simple model of charge-carrier complexes in
coupled quantum-well heterostructures.  We have found that for
indirect trions, the critical layer separation at which the trion
becomes unbound is at least an order of magnitude larger than that of
the biexciton.

We have applied our results to the coupled quantum-well device studied
by Butov \textit{et al.},\cite{Butov2002} as modelled by Tan
\textit{et al.}\cite{Tan2005} We find that, although biexcitons are
unbound in this system, positive and negative trions are bound, with
substantial binding energies.  Qualitatively similar physics is
expected in bilayers of 2D materials, where the interlayer and
intralayer charge-charge interaction potentials reduce to the Coulomb
$1/r$ form studied here at long range. In 2D materials the binding
energy of the trion relative to the biexciton is further magnified by
the nonlocal screening of the charge carriers by the 2D
layers.\cite{Szyniszewski2017}

\begin{acknowledgments}
Computational resources were provided by Lancaster University's
High-End Computing facility. R.\ J.\ H.\ is fully funded by the
Graphene NOWNANO centre for doctoral training (grant
no.\ EP/L01548X/1).  We acknowledge useful conversations with
R.\ J.\ Needs and M.\ Szyniszewski.
\end{acknowledgments}

\bibliography{references}

\end{document}